\pdfoutput=1

\documentclass[aps,prl,twocolumn,superscriptaddress,groupedaddress]{revtex4} %nofootinbib

\usepackage{graphicx}  
\usepackage{dcolumn}   
\usepackage{yfonts}
\usepackage{bm}        
\usepackage{amssymb}   
\usepackage{amsmath}
\usepackage{slashed}
\usepackage{ytableau}
\usepackage{hyperref}
\usepackage[capitalize]{cleveref}

\hypersetup{colorlinks=true,linkcolor=magenta,anchorcolor=green,citecolor=blue,filecolor=black,menucolor=black,urlcolor=blue}

\usepackage{cleveref}
\begin{document}
\widetext

\title{Reconstructing Classical Spacetimes from the S-Matrix in Twistor Space}
% remove these 3 lines before journal submittal.
%\centerline{author list dated 18 June 2018}
% end removal before journal submittal

\author{Alfredo Guevara } \email[]{aguevaragonzalez@fas.harvard.edu}  \affiliation{Center for the Fundamental Laws of Nature, Society of Fellows, \& Black Hole Initiative,
Harvard University, Cambridge, MA 02138, USA}
%
% visitor_addresses.tex                       18 June 2018
%  available symbols are:
%  $\ast, \dag, \ddag, \S, \P, $\|$, $\ast\ast$, \dag\dag, \ddag\ddag ,\#
%
 \noaffiliation
\vskip 0.25cm
      
%\date{}
%%%%%%%%%%%%%%%%%%%%%%%%%%%%%%%%%%%%%%%%%%%%%%%%%%%%%%%%%%%%%%%%%%%%%%%%%%%%%%%%%%%%%%%%%%%%
\begin{abstract}
We present a holographic construction of solutions to the
gravitational wave equation starting from QFT scattering amplitudes. The construction
amounts to a change of basis from momentum to $(2,2)$ twistor space, together
with a recently introduced analytic continuation between $(2,2)$ and $(1,3)$ spacetimes. We
test the transform for three and four-point amplitudes in a classical limit, recovering
both stationary and dynamical solutions in GR as parametrized by their tower
of multipole moments, including the Kerr black hole. As a corollary, this provides a link between the Kerr-Schild classical double copy and the QFT double copy.
\end{abstract}
% \pacs{}
\maketitle
%%%%%%%%%%%%%%%%%%%%%%%%%%%%%%%%%%%%%%%%%%%%%%%%%%%%%%%%%%%%%%%%%%%%%%%%%%%%%%%%%%%%%%%%%%%%%%%%
\section{Introduction}

Can classical spacetimes be understood as fundamental particles? Recently,
a novel body of research has shown that dynamical observables for
compact bodies in GR, such as black holes, can be derived from
massive QFT particles interacting with gravitons, see e.g. \cite{Damour_2016,Cachazo:2017jef,Cheung:2018wkq,Donoghue:1994dn,Bjerrum-Bohr:2018xdl,Jakobsen:2020ksu,Guevara:2017csg,PhysRevD.68.084005,Mougiakakos:2020laz,Arkani-Hamed:2019ymq,Bern:2019nnu,Bjerrum-Bohr:2013bxa,Herrmann:2021tct,Guevara:2018wpp}. This entails that
such spacetimes can be treated as perturbations of flat space sourced by these particles: Perturbative precission is attained by evaluating
the classical limit of scattering amplitudes to the desired order in $G$.  Pushing the state-of-the-art accuracy in GR, the correspondence has mainly emerged through the evaluation of observables such as radiated momentum, waveforms, or on-shell
effective actions \cite{ Cheung:2020gyp,Bern:2019crd,Bern:2021dqo,Cristofoli:2019neg,  Bjerrum-Bohr:2019kec, Bjerrum-Bohr:2021din,Bjerrum-Bohr:2021vuf, DiVecchia:2020ymx,DiVecchia:2021ndb,DiVecchia:2021bdo,Kosower:2018adc, Mogull:2020sak,Kalin:2019rwq,Kalin:2019inp,Levi:2020kvb,Levi:2020uwu,Kalin:2020mvi,Chung:2019yfs,Kalin:2020fhe, Damour:2020tta,Damour:2019lcq, Maybee:2019jus,Bern:2020buy,Chung:2019duq,Chung:2020rrz,Guevara:2019fsj,Aoude:2020onz,Goldberger:2020fot, Goldberger:2017vcg, Goldberger:2017ogt,Chung:2018kqs,Bautista:2019tdr, Bautista:2019evw, Herrmann:2021lqe,Jakobsen:2021smu, Bern:2020uwk,Jakobsen:2021lvp,Mougiakakos:2021ckm,Cristofoli:2020uzm,Vines:2018gqi,Blumlein:2020znm,Carrillo-Gonzalez:2017iyj,Damgaard:2021ipf,Chiodaroli:2021eug,Creci:2021rkz,Damgaard:2021rnk,Gonzo:2021drq,Haddad:2021znf,Kol:2021jjc,Chen:2021qkk}.

The correspondence suggests that asymptotically flat spacetimes can
be reconstructed to some extent from a flat space gravitational S-Matrix. However, even in the classical sense, reconstructing a bulk solution from scattering amplitudes is only natural for the radiative modes reaching null infinity.
This excludes so-called Coulomb or potential modes which can only
explore spatial infinity \footnote{To incorporate Coulomb modes one needs instead to resolve timelike infinity where massive sources localize \cite{Strominger:2017zoo}. We thank A. Strominger for discussions on this.}. To sort this out, in this work we find motivation in the
analytic continuation recently studied in \cite{Crawley}: In $(2,2)$ signature
Coulomb modes can probe null infinity and hence be captured by $(2,2)$
scattering amplitudes. We will provide evidence that these Kleinian spacetimes can indeed be reconstructed from the analytic S-Matrix, after which they can be
Wick rotated back to $(1,3)$ signature. We obtain explicitly stationary and time-dependent solutions even outside their radiation zone, with Schwarzschild and Kerr spacetimes being perhaps our prime applications. Crucially, we rely on the observation that the three- and four-point
S-Matrices for massive particles have a simple classical limit through a multipole expansion
 \cite{Guevara:2018wpp,Bautista:2019evw,Chung:2019duq}. As solutions to the wave equation for compact sources
can be described \textit{exactly} via an infinite tower
of multipoles, we reconstruct them by matching to the multipoles
of the amplitude.

The main geometrical tool of this note is twistor theory. Two complementary perspectives play a role: On the one hand, given that twistors beautifully
parametrize null geodesics, they can be directly used
to reconstruct purely radiative fields from their asymptotic
data at future null infinity, through the Kirchoff-d'Adh\'emar formula \cite{kirchoff,10.2307/2416161,penrose1984spinors,Adamo:2021dfg}. This application emerges naturally
for the $(1,3)$ S-Matrix. On the other hand, twistors were historically used in $(2,2)$ signature to generate exact solutions to the massless equation,
via the Penrose-Ward transform \cite{Tod:1978,Tod:1979tt,Penrose:1976jq,PENROSE197765,Mason:1991rf,WITTEN1978394,Mason:2005qu}. We find that this connects naturally to the $(2,2)$ S-Matrix and enables us to recover even non-radiative modes. As our massive states are classical, the transform only acts on massless particles, in precisely the sense introduced by Witten's twistor string \cite{Witten:2003nn}.

For clarity of exposition we will first be concerned
with a massless scalar, the relation to the gravitational field is explained subsequently.

\section{Twistor Theory}

Let us give a pragmatical introduction to the basic ingredients of twistor
theory, see \cite{White:2020sfn,Chacon:2021hfe,Adamo:2021lrv,Adamo:2021dfg} for recent reviews. As it turns out, the
theory is especially suitable for both $(2,2)$ and $(1,3)$ signatures.
It is then customary to have in mind that the coordinates inhabit a
complefixified spacetime $\{x^{\mu}\}\in\mathbb{C}^{4}$. Now, a twistor
$Z^{\alpha}\in\mathbb{TP}^{4}$ is the projective four-component object
\begin{equation}
Z^{\alpha}=(\lambda^{A},\tilde{\eta}_{\dot{A}})\,\text{\ensuremath{\in\mathbb{C}^{4}}}\,\,\,A,\dot{A}=1,2\,,
\end{equation}
with $\lambda\neq0$. Thus there is a natural fibration $Z^{\alpha}\to\lambda^{A}$
over $\mathbb{CP}^{1}$, which in (1,3) naturally corresponds to the celestial
sphere. More generally, to connect twistor space with flat
space we parametrize the latter by introducing 
\begin{equation}
x_{A\dot{B}}=x_\mu\sigma^\mu_{A\dot{B}}=\left(\begin{array}{cc}
t-x & z-y\\
-z-y & t+x
\end{array}\right)\,,\label{eq:xmain}
\end{equation}
(we use the conventions of App. E of \cite{Crawley}; both $x_{A\dot{B}}$ and $Z^\alpha$ will be real in $(2,2)$ signature). Then, the condition
\begin{equation}
\rho:\,\,\tilde{\eta}^{\dot{A}}=x^{\dot{A}B}\lambda_{B}\,,\label{eq:incf}
\end{equation}
is known as the \textit{incidence relation}. Through it, twistors parametrize null geodesics: For a fixed $Z^{\alpha}$ the
set of points satisfying this condition form a null ray.

We will use $\rho_{x}[f(Z^{\alpha})]$ to mean that we evaluate $f(Z^{\alpha})$
on the support of the incidence map. Consider now a holomorphic function
$f_{-2}(Z^{\alpha})$ which will encode some dynamical data. For a closed contour $\mathcal{C}$ in the Riemann sphere of $\lambda^A$, the \textit{Penrose transform} is
\begin{equation}
\phi_{f}(x)=\int_{\mathcal{C}}\frac{\langle\lambda d\lambda\rangle}{2\pi}\times\rho_{x}[f_{-2}(Z^{\alpha})]\,.\label{eq:sca}
\end{equation}
It is essentially an integral over null directions passing through
$x$. The notation $f_{-2}$ means the function is homogeneous of degree
$-2$, as required by the identification $Z^{\alpha}\sim tZ^{\alpha}$.
In practice, we can fix the scale by taking $\lambda^{A}=(1\,\,z)$.

We shall assume that $f_{-2}(Z)$ has finite many poles, encoding physical data,
and that the contour $\mathcal{C}$ defines two hemispheres that contain at least one pole each. The Penrose transform then constructs non-trivial exact solutions of the wave equation: As the function $f_{-2}$ is holomorphic, i.e. independent of $\bar{Z}$, we have that
\begin{equation}
\frac{\partial}{\partial x^{\dot{A}A}}\rho_{x}[f(Z^{\alpha})]=\lambda_{A}\rho_{x}[\frac{\partial f(Z^{\alpha})}{\partial\tilde{\eta}^{\dot{A}}}]\,,
\end{equation}
from which it follows immediately that (\ref{eq:sca}) satisfies
\begin{equation}
\partial_{\dot{A}A}\partial^{A\dot{A}}\phi_{f}=0\,\,\,\,.
\end{equation}
Even though the field appears free, singularities of the integral  (\ref{eq:sca}) in $x^\mu$ encode compact sources when interpreted in $(3,1)$ signature. We anticipate that $\phi_{f}$ will be identified with gravitational
perturbations as in \cite{Crawley}.

\section{Analytic Continuation And Reconstruction}

We initiate our discussion of physics in $(1,3)$ before continuing
to $(2,2)$. The analytic continuation picture will suggest that a
twistor S-Matrix based on Feynman (F) rather
than retarded (R) propagators can indeed be used to fully reconstruct
exact solutions. To understand the difference we consider both cases
simultaneously, using $c=F,R$ respectively.

We start with perturbations in flat space, as described by a scalar
$\phi(x)$ sourced by an effective current $\mathcal{J}(k)$. It shall
become clear that the scalar is related to a gravitational perturbation to a desired accuracy in $G$.
The two aforementioned solutions are
\begin{equation}
\Phi_{c}(x)=\int\frac{d^{4}k}{(2\pi)^{4}}e^{ikx}\frac{\mathcal{J}(k)}{(k^{0})^{2}-\vec{k}^{2}+i\mu(c)}\,,c=F,R\label{eq:adv}
\end{equation}
which solely differ by the $i\epsilon$ prescription, $\mu(F)=\epsilon$
whereas $\mu(R)=\epsilon k^{0}$. They correspond to different
boundary conditions, in the Feynman case the result is a time-symmetric
solution and hence boundary conditions are required at both past and
future null infinity. 

Both fields contain Coulomb or stationary modes $k^{0}\approx0$ directed
towards spatial infinity. On the other hand, the modes directed towards
null infinity have $k^{2}=0$ and correspond to radiation/free modes.
In $(1,3)$ such momentum regions do not overlap reflecting that stationary
solutions do not radiate. Now, to connect to the S-Matrix we first
aim to extract the radiative modes from the field. It turns out that
this can be done easily if we parametrize the integration via spinors
\begin{equation}
k_{A\dot{A}}=\omega\lambda_{A}\tilde{\lambda}_{\dot{A}}+\xi q_{A\dot{A}}\,,\label{eq:kdc}
\end{equation}
where $|\lambda\rangle,[\tilde{\lambda}|$ are conjugate coordinates
on $\mathbb{CP}^{1}$, which we shall think of as a celestial sphere.
Also $\omega,\xi\in\mathbb{R}$ and $q$ is a null reference vector,
which we can take such that $\langle\lambda|q|\tilde{\lambda}]>0$.
After some computation we obtain
\begin{align}
\Phi_{c}(x) {=}&{\int_{\mathbb{CP}^{1}}}\frac{\langle\lambda d\lambda\rangle[\tilde{\lambda}d\tilde{\lambda}]}{(2\pi)^{2}}\int_{\mathbb{R}}\frac{\omega d\omega e^{i\frac{\omega}{2}\langle\lambda|x|\tilde{\lambda}]}}{4(2\pi)^{2}}\int_{\mathbb{R}}\frac{d\xi e^{i\xi q\cdot x}}{\xi{-}i\bar{\mu}(c)}\mathcal{J}(k)\,,\nonumber \\
 & \bar{\mu}(F)=\epsilon\omega\,\,\,,\bar{\mu}(R)=\epsilon\,,\label{eq:ieps}
\end{align}
The way to extract the radiation modes with $k^{2}=0$ is now clear:
We simply pick up the pole at $\xi=0$. Thus we drop, in principle,
singularities in the source $\mathcal{J}(k(\xi))$ which may lead to extra (e.g.
Coulomb) modes. But we will see that such information is still there. 

Assuming w.l.o.g. that $q\cdot x>0$ we can close the $\xi$ contour
upwards. Crucially, because of different $i\epsilon$ prescriptions in
\eqref{eq:ieps}, we note that in the case $c{=}F$ the pole at $\xi{=}0$
only contributes for $\omega>0$, i.e. the integral is restricted
to positive frequencies. The two results are
\begin{align}
\phi_{c}(x)=\int_{\mathbb{CP}^{1}}&\frac{\langle\lambda d\lambda\rangle[\tilde{\lambda}d\tilde{\lambda}]}{(2\pi)^{2}}\int_{L(c)}^{\infty}\frac{i\omega d\omega e^{i\frac{\omega}{2}\langle\lambda|x|\tilde{\lambda}]}}{4(2\pi)}\mathcal{J}(k)\,,\nonumber\\
&L(F)=0\,\,,\,\,\,L(R)=-\infty\,.\label{eq:phioc}
\end{align}
In fact, we have just recovered the textbook wisdom that the Feynman
propagator only incorporates a particle $\omega>0$, whereas the retarded
case incorporates both particle $(\omega>0)$ and antiparticle $(\omega<0$).
By performing the full integral over $\omega\in\mathbb{R}$ the Coulomb
modes cancel out between particle/antiparticle; such a signal only
propagates in the light cone. In contrast, by restricting to $\omega>0$,
i.e. the Feynman prescription, the modes indeed have support outside
the light-cone; hence there is hope to recover Coulomb components
from amplitudes! 

To interpret the formula \eqref{eq:phioc} consider first $c{=}R$. We then identify
the $\omega$ integral as $\psi(\ell{\cdot} x,\ell^{\mu})$,
where the radiative data at null infinity is described by
\begin{equation}
\psi(u,\ell^{\mu}):=\frac{1}{8\pi^2}\int_{-\infty}^{\infty}i\omega d\omega e^{i\omega u}\mathcal{J}(\omega\ell)\,,
\end{equation}
see appendix I. Here $\ell=|\lambda\rangle[\tilde{\lambda}|$ parametrizes the celestial
sphere and $u$ is a retarded time. In this case, our equation \eqref{eq:phioc}
is nothing but a (free) bulk field reconstructed from its radiative
data $\psi$, by inverting the retarded propagation. Now, quite remarkably, it has been argued that the modes
of $\psi(u)$, being purely on-shell, are given by a coherent-classical
limit of scattering amplitudes \cite{Monteiro:2020plf,Cristofoli:2021vyo,Bautista:2021wfy,Bautista:2021llr}. For a massless particle of momentum $\omega|\lambda\rangle[\tilde\lambda| $,
\begin{equation}
\mathcal{J}(k)=\mathcal{M}(\omega|\lambda\rangle[\tilde\lambda|)\,\,\,\:\textrm{at }\,\,\,k^{2}=0\,\,,\label{eq:jfs}
\end{equation}
(other particles are implicit in the RHS). Now, what about the other case $c{=}F$? It does not yield such a simple interpretation in $(1,3)$ signature. It turns out, the picture becomes strikingly clear in $(2,2)$ where the scattering
amplitudes have natural support. Let us then continue \eqref{eq:phioc} and \eqref{eq:jfs}
by imposing that the product $\langle\lambda|x|\tilde{\lambda}]$ is
invariant in form; this entails that $\langle\lambda d\lambda\rangle[\tilde{\lambda}d\tilde{\lambda}]\rightsquigarrow i\langle\lambda d\lambda\rangle[\tilde{\lambda}d\tilde{\lambda}]$
and that $|\lambda\rangle$ and $[\tilde{\lambda}|$ are real
and independent, i.e. $\mathbb{CP}^{1}\rightsquigarrow\mathbb{R}\times\mathbb{R}$. Introducing $|\tilde{\mu}]:=\omega|\tilde{\lambda}]$ we get
\begin{align}
\phi_{c}(x)=&\int_{\mathbb{R}}\frac{\langle\lambda d\lambda\rangle}{2\pi}\,\rho_{x} \mathfrak{M}^{(c)}(\lambda,\tilde{\eta})\,\,\,\,\textrm{with} \nonumber\\
\mathfrak{M}^{(c)}(\lambda,\tilde{\eta})
=& \int_{\mathbb{R}{\times} L(c)} \frac{d^2\tilde\mu}{(4\pi)^2}\,e^{i[\tilde{\eta}\tilde{\mu}]/2}\,\mathcal{M}(|\lambda\rangle[\tilde\mu|)\,\,\,\,\,,
\label{eq:fm2}
\end{align}
where the first line is precisely a Penrose transform \eqref{eq:sca}!
The relation to $\mathfrak{M}^{(c)}$, the classical amplitude in $(2,2)$ twistor space,
constitutes the main result of this note. For $c=R$ the integration covers the full $\tilde\eta$ plane; this is the usual definition of the twistor S-Matrix in the context of $\mathcal{N}=4$ SYM \cite{Witten:2003nn,Arkani-Hamed:2009hub,Mason:2009sa}. In contrast, the Feynman case yields a new transform that excludes the `antiparticles' \footnote{There is no intrinsic definition of antiparticles in $(2,2)$ as $\textrm{sgn}(k^0)$ is not $sl(2,\mathbb{R})$ invariant (this may be related to the discussion in \cite{Arkani-Hamed:2009hub}). This is the one of the reasons we resort to analytic continuation.} and captures the Coulomb modes, where $|\tilde\mu]=\omega (1\,\,\ z)$ and $\omega>0$. Here the energy integration simply becomes a Laplace transform.

Next we provide applications of the new formula for both stationary
and non-stationary cases. The usual downside of the Feynman prescription, namely that it requires boundary data at both past and future null infinity,
disappears in this case since $(2,2)$ Kleinian space has only a single null infinity \cite{Crawley}.

\section{Three-point Amplitudes: Stationary Solutions}

We have recently argued that linearized stationary black holes can be decomposed purely in terms of on-shell modes when continued to $(2,2)$ signature \cite{Crawley}. Although implicit, the
continuation done there is equivalent to the Feynman prescription \footnote{The related work \cite{Monteiro:2020plf} has examined the retarded
$(2,2)$ propagators obtained from the KMO formalism \cite{Kosower:2018adc}.}. We shall see that these results also emerge naturally from twistor three-point amplitudes (see appendix IV for comparison). 

Consider a $p_{1}\to k+p_{2}$ process involving a massive particle,
representing the source, and massless particle of momentum $k^{\mu}$.
The process is classical if the massive particles are very heavy.
Thus the massive momenta defines a classical time direction via
\begin{equation}
p_{1}^{\mu}\approx p_{2}^{\mu}\approx Mu^{\mu}\,,
\end{equation}
Momentum conservation yields $p_{2}^{2}-p_{1}^{2}\approx2Mu\cdot k$
so that the corresponding classical source must have the form
\begin{equation}
\mathcal{J}(k)=2\pi\delta(u\cdot k)j(k)\,,\label{eq:stcr}
\end{equation}
which means it is time independent and associated to Coulomb modes \footnote{The function can be regarded as analytic by regulating the Coulomb pole as $\frac{1}{u\cdot k + i\epsilon}- \frac{1}{u\cdot k - i\epsilon}$.} .
To connect this to a three-point S-Matrix we need to make sure that
its kinematics, i.e. $u\cdot k=k^{2}=0$, have solutions. As anticipated this is impossible in $(1,3)$ signature, but
happens naturally in $(2,2)$ signature! Plugging \eqref{eq:stcr} into \eqref{eq:fm2} localizes $\tilde\mu^{\dot {A}} \to \omega\, u^{\dot{A} B} \lambda_B $, giving
\begin{equation}
\mathfrak{M}_3(\lambda,\tilde{\eta}){=}\int_{0}^{\infty}\frac{d\omega}{4\pi}e^{-\frac{\omega}{2}[\tilde{\eta}|u|\lambda\rangle}M_{3}(\omega|\lambda\rangle\langle\lambda|u)\,,\,c{=}F,\label{eq:fm2m3}
\end{equation}
where $M_{3}(k)=j(k)$ at $k^2=0$, while $c=R$ will lead to the anticipated vanishing result \footnote{Hereafter we rotate $x^{\mu}\to ix^{\mu}$,
which preserves the $(2,2)$ signature as explained in appendix IV. Through the incidence relation
\eqref{eq:incf} this yields $[\tilde{\mu}|\to i[\tilde{\mu}$. In
turn, the spinor $x_{A\dot{B}}$ is now given by \eqref{eq:xmain}
with real components.}. Now, what is the most general form of a stationary current? In QFT the
currents for massless quanta are described in terms of form factors and multipoles. General multipoles do not have an intrinsic definition, having
been developed via many different approaches both on QFT and
classical GR fronts \cite{Cotogno:2019vjb,Lorce:2009br,Lorce:2019sbq,doi:10.1063/1.1666501,osti_5402287,
doi:10.1063/1.1665427,doi:10.1063/1.1704349,Compere:2017wrj,Blanchet:2013haa}. Luckily, for our amplitudes there is an intrinsic definition:
For the massive particle, the spin degrees of freedom lead to a classical
Pauli-Lubanski vector $a^{\mu}$ \cite{Bautista:2019tdr,Chung:2019duq,Guevara:2018wpp,Maybee:2019jus,Aoude:2021oqj}, which
satisfies
\begin{equation}
a\cdot u=0\,.
\end{equation}
It then essentially follows from its on-shell conditions that the general amplitude becomes, in the classical limit,
\begin{equation}
M_{3}(k)=\sum_{n}\frac{c_{n}}{n!}(a\cdot k)^{n}\,,\label{eq:m3mp}
\end{equation}
for constants $c_{n}$. We derive and elaborate on this result for our scalar mode in Appendices II and III, both in vector and twistor pictures.
The crucial take-away is that the operators $k^{\mu_{1}}\cdots k^{\mu_{n}}$ are symmetric trace-free (STF) and transverse due to the conditions $u\cdot k{=}k^{2}{=}0$.
Thus they project the operators $a^{\mu_{1}}\cdots a^{\mu_{n}}$ into
their STF piece, i.e. irreps of
the massive little-group $sl(2,\mathbb{R})$. This
is precisely the defining property of the multipole expansion, which
here just follows from the on-shell kinematics of the $(2,2)$ three-point
amplitude.

Indeed, the above becomes explicit in converting \eqref{eq:m3mp} to twistor space. Inserting it in (\ref{eq:fm2m3}) (see appendix IV for convergence) the result is
\begin{equation}
\mathfrak{M}_3(|\lambda\rangle,x|\lambda\rangle)=\sum_{n}\frac{c_{n}}{2\pi}\frac{\langle\lambda|A|\lambda\rangle^{n}}{\langle\lambda|X|\lambda\rangle^{n+1}}\,,\label{eq:dfd}
\end{equation}
where we have introduced the spatial projections
\begin{equation}
X_{AB}=x_{(A\dot{B}}u_{\,\,B)}^{\dot{B}}\,\,\,,\,A_{AB}=a_{(A\dot{B}}u_{\,\,B)}^{\dot{B}}\,\,,\label{eq:key}
\end{equation}
corresponding to $x_{\mu}u_{\nu}\sigma^{\mu\nu}$ and $a_{\mu}u_{\nu}\sigma^{\mu\nu}$. From \eqref{eq:xmain} we note that we can define spatial distance
as
\begin{equation}
r^{2}:=\det(X)=z^{2}-x^{2}-y^{2}\,.\label{eq:rfx}
\end{equation}
Feeding this into the Penrose transform (\ref{eq:sca}), using \eqref{eq:ide},
gives the sought classical solution
\begin{align}
\phi(x) & =\frac{1}{4\pi}\sum_{n}\frac{(-1)^{n}}{n!}Q^{A_{1}{\cdots} A_{2n}}\frac{\partial}{\partial X^{A_{1}A_{2}}}{\cdots}\frac{\partial}{\partial X^{A_{2n-1}A_{2n}}}\frac{1}{r}\,,\label{eq:multpl}
\end{align}
where the multipole moments are $Q^{A_{1}\cdots A_{2n}}=c_{n}A^{(A_{1}A_{2}}\cdots A^{A_{2n-1}A_{2n})}$
in agreement with the twistor multipoles of \cite{Chacon:2021hfe,curtis}. Here
we have used the residue theorem
\begin{equation}
\int_{\mathbb{R}}\frac{1}{2\pi}\frac{\langle\lambda d\lambda\rangle}{\langle\lambda|X|\lambda\rangle}=\frac{1}{2\sqrt{\det(X)}}=\frac{1}{2r}\,,
\end{equation}
by noticing that the contour $\mathcal{C}=\mathbb{R}$ separates the
two conjugate roots of $\langle\lambda|X|\lambda\rangle$, as required
for the Penrose transform. An advantage of this language is that the implicit STF tensors have explicitly become the fully symmetric $sl(2,\mathbb{R})$ irreps appearing in \eqref{eq:multpl} or \eqref{eq:dfd}. They correspond to $m=0$ harmonics as shown in appendix III.

To rotate back to $(1,3)$ signature we can use $(x,y)\to i(x,y)$
in \eqref{eq:rfx}. We see that the terms in \eqref{eq:multpl}
scale as $1/r^{n}$, and are thus hidden from the $1/r$
order that defines radiation at null infinity. This reveals that we
have recovered the full tower of Coulomb modes and that the field is \textit{not} free in $(1,3)$. A perhaps more familiar
form of \eqref{eq:multpl} is 
\begin{equation}
\phi(x)=\sum_{n}\frac{(-1)^{n}}{4\pi}Q^{i_{1}\cdots i_{n}}\times\frac{\partial}{\partial x^{i_{1}}}...\frac{\partial}{\partial x^{i_{n}}}\left(\frac{1}{r}\right)\,,
\end{equation}
where $Q^{i_{1}\cdots i_{n}}=c_{n}a^{i_{1}}\cdots a^{i_{n}}$ as expected.
This is the multipole form of the general stationary solution for compact sources. It is remarkable that we have reconstructed it from an a priori radiative
three-point S-Matrix. Indeed, the key ingredients in the characterization
of the solution, given in (\ref{eq:key}), are here directly
obtained from linear and angular momentum of QFT massive particles.

\section{Gravitational Field from a Scalar Potential}

It is easy to extend the above to gravity or gauge theory and connect
it with the Penrose transform for higher helicity. For a vacuum
solution the gravitational field is naturally associated to higher multipole
moments e.g. \cite{Chacon:2021hfe,curtis}.
Too see how this relation emerges here, consider first a linearized $O(G)$
metric $h_{\mu\nu}=g_{\mu\nu}-\eta_{\mu\nu}$ that is stationary with
respect to a time direction $u^{\mu}$.

Following \cite{Crawley}, we can decompose it into (anti)self-dual components
$h_{\mu\nu}^{+}$ and $h_{\mu\nu}^{-}$, and then consider the scalar
function
\begin{equation}
\phi^{\pm}=u^{\mu}u^{\nu}h_{\mu\nu}^{\pm}=h_{00}^{\pm}\,.
\end{equation}
Being the norm of a Killing direction, $\phi^{\pm}$ are scalar invariants
in the linearized theory. In fact, these two scalars are nothing but
the two on-shell degrees of freedom of a free massless particle as explained in appendix II.

The two self-dual pieces of the curvature tensor can be obtained in spinor
components. Recalling that $\sigma_{AB}^{\mu\nu}$ is self-dual in
its Lorentz indices, the projection is given by
\begin{align}
C_{ABCD}^{+} & =2\sigma_{AB}^{0i}\sigma_{CD}^{0j}\partial_{i}\partial_{j}h_{00}^{+}\nonumber \\
 & =2\frac{\partial}{\partial X^{AB}}\frac{\partial}{\partial X^{CD}}\phi^{+}(x)\,,\label{eq:phipp}
\end{align}
where $\sigma_{AB}^{0i}x_{i}=\frac{1}{2}X_{AB}$, $\sigma_{0i}^{AB}\partial_{AB}=\frac{1}{2}\partial_{i}$.
The conjugate relation holds for $C_{\dot{A}\dot{B}\dot{C}\dot{D}}^{-}$.
Now, we argued in \cite{Crawley} that $\phi^{\pm}$ are in correspondence
with two 3-point amplitudes $M_{3}^{\pm}$, representing graviton emission
from a spinning particle. Using this, we can insert the Penrose
transform (\ref{eq:fm2}),(\ref{eq:fm2m3}) into equation (\ref{eq:phipp}), leading to the formulae
\begin{align}
C_{ABCD}^{+} & =\int_{\mathcal{C}}\frac{\langle\lambda d\lambda\rangle}{2\pi}\times\lambda_{A}\lambda_{B}\lambda_{C}\lambda_{D}\rho_{x}[\mathfrak{M}_3^{+}(Z^{\alpha})]\,,\label{eq:pr1}\\
C_{\dot{A}\dot{B}\dot{C}\dot{D}}^{-} & =\int_{\mathcal{C}}\frac{\langle\lambda d\lambda\rangle}{2\pi}\times\rho_{x}[\frac{\partial^{4}}{\partial\tilde{\eta}^{\dot{A}}\partial\tilde{\eta}^{\dot{B}}\partial\tilde{\eta}^{\dot{C}}\partial\tilde{\eta}^{\dot{D}}}\mathfrak{M}_3^{-}(Z^{\alpha})]\,,\label{eq:pr2}
\end{align}
where we have followed analogous steps for $\phi^{-}$. The homogeneous
functions (of degrees $-6$ and $+2$) are 
\begin{align}
\mathfrak{M}_3^{+}(Z^{\alpha}) & =\frac{1}{4}\int_{0}^{\infty}\frac{d\omega}{2\pi}\omega^{2}e^{-\frac{\omega}{2}[\tilde{\eta}|u|\lambda\rangle}M_{3}^{+}(\sqrt{\omega}|\lambda\rangle,\sqrt{\omega}\langle\lambda|u)\label{eq:gp1}\\
\mathfrak{M}_3^{-}(Z^{\alpha}) & =4\int_{\varepsilon\to 0}^{\infty}\frac{d\omega}{2\pi}\frac{1}{\omega^{2}}e^{-\frac{\omega}{2}[\tilde{\eta}|u|\lambda\rangle}M_{3}^{-}(\sqrt{\omega}|\lambda\rangle,\sqrt{\omega}\langle\lambda|u)\label{eq:gp2}
\end{align}
(the IR regulator $\varepsilon$ does not contribute to \eqref{eq:pr2}). Formulae (\ref{eq:pr1})-(\ref{eq:pr2}) are indeed the Penrose transform
for the curvature tensor \cite{Vinesx,penrose1984spinors}. Here we have connected
it to three-point amplitudes through (\ref{eq:gp1}) and (\ref{eq:gp2}).
Similar formulae hold at higher orders in $G$, as well as non-stationary metrics as detailed
somewhere else \cite{Bautista2}.

To present a direct application of the construction, let us focus on the Kerr black hole and compute its curvature tensors (\ref{eq:pr1})-(\ref{eq:pr2}). We start with the leading order in $G$: Remarkably,
it corresponds to the \textit{minimally coupled} 3-point amplitudes introduced in \cite{Guevara:2017csg,Guevara:2018wpp,Chung:2018kqs}. They can be obtained from our generic multipole expansion (\ref{eq:m3mp})
by fixing the coefficients as $c_{n}^{\pm}=4\pi GM(\pm1)^{n}$, i.e.
\begin{equation}
M_{3}^{\pm}(\sqrt{\omega}|\lambda\rangle,\sqrt{\omega}\langle\lambda|u)=4\pi GMe^{\mp \frac{\omega}{2}\langle\lambda|A|\lambda\rangle}\,,\label{eq:3ptke}
\end{equation}
where we used (\ref{eq:key}). Inserting this into (\ref{eq:gp1}) we get
\begin{align}
\mathfrak{M}_3^+(|\lambda\rangle,x|\lambda\rangle) & {=}\frac{GM}{2}\int_{0}^{\infty}d\omega\omega^{2}e^{-\frac{\omega}{2}\langle\lambda|J_{+}|\lambda\rangle}{=}\frac{8GM}{\langle\lambda|J_{+}|\lambda\rangle^{3}}\,.\label{eq:lJl}
\end{align}
whereas $\mathfrak{M}_3^{-}(Z^{\alpha})$ is a function of $J_{AB}^-$. Here
\begin{equation}
J_{AB}^{\pm}=X_{AB}\pm A_{AB}=2(x_{[\mu}u_{\nu]}\pm a_{[\mu}u_{\nu]})\sigma_{AB}^{\mu\nu}\,.\label{eq:gde}
\end{equation}
We recognize
them as the angular momentum and Killing-Yano tensors of the Kerr metric \footnote{They are given by $J=J_{AB}^{-}\epsilon_{\dot{A}\dot{B}}+\tilde{J}_{AB}^{+}\epsilon_{AB}$
and $Y=J_{AB}^{+}\epsilon_{\dot{A}\dot{B}}+\tilde{J}_{AB}^{-}\epsilon_{AB}$
where $\tilde{J}_{\dot{A}\dot{B}}^{\pm}=\tilde{u}_{\dot{A}}^{\,\,A}\tilde{u}_{\dot{B}}^{\,\,B}J_{AB}^{\pm}$ \cite{penrose1984spinors,Vinesx}.}. The relative
sign signals that the spin $a_{\mu}$ is a pseudovector. They have an important geometric meaning: The Penrose
transform of \eqref{eq:lJl} is a simple quadrupolar integral 
\eqref{eq:exint} and gives
\begin{equation}
C_{ABCD}^{+}=\frac{3 GM/2}{\left[(z+a)^{2}-x^{2}-y^{2}\right]^{5/2}}J_{(AB}^{+}J_{CD)}^{+}\,.\label{eq:JJA}
\end{equation}
Now, any symmetric spinor can be written in terms of its roots $J_{AB}^{+}{=}\mu_{(A}\kappa_{B)}$ satisfying $\langle \mu \kappa\rangle^2{=}-4|J^+|$ \footnote{This follows from writing the quadratic form $\langle\lambda|J_{+}|\lambda\rangle$ as $\langle\lambda\mu\rangle\langle\kappa\lambda\rangle$ for all $\lambda$. The roots are conjugate as required by the evaluation of \eqref{eq:exint}.}. We normalize them by introducing $|\hat\kappa\rangle := |\kappa\rangle/\langle\mu\kappa\rangle$.
Further continuing the coordinates to $(1,3)$ signature as in \cite{Crawley},
we obtain
\begin{equation}
C_{ABCD}^{+(1,3)}=- \frac{6 GM}{(r+ia\cos\theta)^{3}}\mu_{(A}\hat\kappa_{B}\mu_{C}\hat\kappa_{D)}\,,\label{eq:exac}
\end{equation}
i.e. the known form of the curvature tensor of Kerr \cite{kinnersley}. The
decay $\sim1/r^{3}$ again confirms Coulomb modes well hidden from the usual
radiative expansion. Moreover, the Penrose transform has given us an exact result in $G$: This will occur for the classical spacetimes satisfying the so-called Kerr-Schild/Weyl double copy of \cite{Luna:2018dpt,Chacon:2021wbr}.
By definition, the spinors $\mu_{A}$ and $\hat\kappa_{A}$
form the principal null directions (PNDs) of the spacetime. Even more,  (\ref{eq:gde}) is nothing but the sum of orbital and
intrinsic angular momenta of the Kerr black hole \cite{penrose1984spinors}!
It is remarkable that here it originates directly from the QFT momentum
and spin vector of massive particles.

\section{Four-point amplitudes: Radiative Moments}\label{sec:4pt}

As a last example we consider time-dependendent perturbations of spacetimes, which emerge naturally in their dynamics. One can for instance consider a scalar perturbation
$\phi$ of a Schwarzschild background controlled by the Regge-Wheeler
equation; such perturbations impinge the source from past null infinity
and radiate to future null infinity \cite{futterman_handler_matzner_1988}. The extension to Gravitational Waves is straightforward. The QFT description of
such processes in terms of a four-point S-Matrix in $(1,3)$ was
recently studied in \cite{Bautista:2021wfy,Bautista2}. 

To see how this emerges here, we extend the previous case by considering
an additional wave of momentum $k'$ scattering in the source, i.e.
$p_{1}+k'\to p_{2}+k$. The corresponding on-shell condition is $p_{2}^{2}-p_{1}^{2}\approx2Mu\cdot(k-k')$
and hence
\begin{equation}
\mathcal{J}(k)=2\pi\delta(u\cdot k-u\cdot k')j(k)\label{eq:jkg}
\end{equation}
In $(1,3)$ signature, this is radiative because the massive source
is accelerated by an incoming wave. Identifying the source with a
linearized black hole, it was found in \cite{Bautista:2021wfy} that the Regge-Wheeler
amplitudes $j(k)=M_{4}(k)$ solely depend on the direction of $k^{\mu}=\omega\ell^{\mu}$
in the long-wavelength regime. Thus they can be easily
written in terms of spherical harmonics, see appendix III,
\begin{align}
M_{4}(k) & =m_{0}+m_{1}\mathcal{\ell\cdot}d/\ell\cdot u+\ldots\nonumber \\
 & =m_{0}+m_{1}\frac{\langle\lambda|d|\tilde{\lambda}]}{\langle\lambda|u|\tilde{\lambda}]}+\ldots\label{eq:dipm4}
\end{align}
where the dipole vector satisfies $d\cdot u=0$ and we have parametrized
$\ell=|\lambda\rangle[\tilde\lambda|$.
The numerical moments $\{m_{i}\}$ can be computed from QFT.

We can reconstruct a bulk solution from the radiative Regge-Wheeler
amplitude by proceding via $(2,2)$ signature again. We illustrate
this with the two terms in \eqref{eq:dipm4}. Plugging \eqref{eq:jkg}
into the Feynman Twistor transform (\ref{eq:fm2}), we find  $\mathfrak{M}_4(|\lambda\rangle,x|\lambda\rangle)$ to be given by
\begin{equation}
\frac{e^{-\omega'\frac{\langle\lambda|xu|\xi\rangle}{\langle\lambda\xi\rangle}}}{2\pi}\left(\frac{m_{0}}{\langle\lambda|X|\lambda\rangle}{+}\frac{im_{1}}{\omega'}\frac{\langle\lambda|D|\lambda\rangle}{\langle\lambda|X|\lambda\rangle^{2}}{+}im_{1}\frac{\langle\lambda|D|\xi\rangle}{\langle\lambda\xi\rangle\langle\lambda|X|\lambda\rangle}\right)
\end{equation}
where $\omega'=u\cdot k'$ and $D_{AB}=d_{(A\dot{B}}u_{\,\,B)}^{\dot{B}}$.
Here $|\xi\rangle$ is a reference spinor. The first two terms are
again Coulomb multipoles in twistor space, c.f. \eqref{eq:dfd}. The
last term is a new, radiative contribution $\sim1/r$. The solution
is time dependent, as can be seen by using (\ref{eq:key}) to decompose
$x_{A\dot{A}}u_{\,\,B}^{\dot{A}}=X_{AB}+\epsilon_{AB}t$ in the exponent.
Plugging this into the Penrose transform, analogously to \eqref{eq:multpl},
we obtain
\begin{equation}
\phi_{\omega'}=\frac{e^{-i\omega'(r-it)}}{4\pi}\left(\frac{m_{0}}{r}-\frac{im_{1}}{\omega'}d\cdot\hat{x}\left[\frac{1}{r^{2}}+i\omega'\frac{1}{r}\right]\right)\,.\label{eq:phifd}
\end{equation}
We have recovered an exact solution of the $(2,2)$ wave equation,
including both Coulomb and radiative modes. At face value it is neither
an `in' or `out' solution but resembles a quasinormal mode decaying
in time. However, one can obtain a $(1,3)$ solution simply by rotating
$t\to-it$. It corresponds to an outgoing mode, which at null infinity
yields the radiative data
\begin{equation}
\lim_{r\to\infty}r\phi_{\omega'}=\frac{e^{i\omega'u}}{4\pi }\left(m_{0}+m_{1}\frac{d\cdot\ell}{u\cdot\ell}\right)\,.
\end{equation}
Here the retarded time $u:t-r$ is fixed and $\ell$ now corresponds to the celestial sphere. As in \cite{Bautista:2021wfy}, this translates directly into our starting 
data \eqref{eq:dipm4}, see also appendix I. The interpretation of \eqref{eq:phifd}
is now clear: It is the outgoing mode associated to a monopole-dipole
excitation of frequency $\omega'$. Indeed, smearing over $\omega'$
frequencies yields a more familiar form of such solution:
\begin{equation}
\int d\omega'\tilde{a}(\omega')\partial_{u}\phi_{\omega'}=\frac{m_{0}\partial_u a(u)}{r}+m_{1}d\cdot\hat{x}\left[\frac{a(u)}{r^{2}}+\frac{\partial_u a(u)}{r}\right]\,\,\,\,
\end{equation}
where $\tilde{a}(\omega')$ and $a(u)$ are Fourier conjugates. Finally
note that in deriving the form \eqref{eq:jkg} we have also assumed
that an incoming solution of frequency $\omega'$ exists. In $(1,3)$
the incoming solution is needed by physical considerations to cancel
out outgoing radiation from the black hole \cite{futterman_handler_matzner_1988}. However, from the perspective
of the $(2,2)$ space, the sole outgoing solution appears consistent
since there is no black hole source at $r=0$ \cite{Crawley}.

\section{Open questions}

Via twistor theory, we have just scratched the surface of using the analytic S-Matrix to reconstruct solutions of gravitational
wave equations. Remarkably, both radiative and Coulomb modes arise on the same footing in $(2,2)$, and in a sense correspond to purely outgoing plane waves. In turn, from our last discussion it would seem that we lost the information of $(1,3)$ incoming radiation, although it may be possible to use crossing symmetry
of the S-Matrix to recover it.

On the other hand, the fact that null
infinity has only one connected component in $(2,2)$ strongly suggests
that the reconstruction is suitable for realistic scenarios with no-incoming radiation \cite{maggiore2018gravitational,Blanchet:2013haa}, e.g. the generation of gravitational waves. The associated
amplitudes may have more matter sources but only one outgoing graviton \cite{Bautista:2019evw,Carrasco:2020ywq,Carrasco:2021bmu,Jakobsen:2020ksu,Jakobsen:2021lvp}, making them suitable for the techniques presented (see \cite{Damour:2020tta} for a recent discussion of retarded vs Feynman propagators in this context). They are controlled to some extent by soft factorization \cite{Bautista:2019tdr}: Indeed the amplitude $M_3$ treated here is the simplest instance of such, where the Coulomb poles of the soft factor $\frac{1}{u\cdot k \pm i\epsilon}$ are equivalent to the delta function $\delta(u\cdot k)$.

Our twistor transform \eqref{eq:fm2} amounts to a change of basis of the gravitational S-Matrix. Similar transforms have unveiled new symmetries of the gauge-theory S-Matrix to all loop orders, see e.g. \cite{Hodges:2009hk,Elvang:2013cua}. It would be interesting to explore this vein for our amplitudes. Indeed, our construction relies
on the relation between classical spacetimes and gravitational amplitudes, which has been tested to high loop precision \cite{Bautista:2021llr,Bern:2021dqo,Bern:2020uwk,Jakobsen:2021zvh,Kalin:2020fhe}. It is a pressing question to understand how non-linear effects interrelate both sides, especially in view that the gravitational field is not free asymptotically in the non-linear theory, which is reflected as an IR dressing in the perturbative theory. 

For certain scenarios such as Kerr-Schild (KS) spacetimes the linearized examples we have discussed should suffice. In fact, \cite{White:2020sfn,Chacon:2021hfe} have argued that the Kerr-Schild double copy structure \cite{Monteiro:2014cda,Luna:2015paa,Luna:2016due,Luna:2016hge,Godazgar:2020zbv,Godazgar:2021iae,Luna:2018dpt,Easson:2021asd,Chacon:2021wbr,Mao:2021kxq} follows from the Penrose transform. Due to this, it can be readily checked that our formula \eqref{eq:fm2} links the KS double copy directly to the double copy between gauge and gravity massive amplitudes of  \cite{Bautista:2019tdr}, also developed in  \cite{Bautista:2019evw,Brandhuber:2021bsf,Li:2021yfk,Brandhuber:2021eyq,Carrasco:2021bmu,Gonzalez:2021bes,Johansson:2019dnu,Momeni:2020hmc,Moynihan:2021rwh,Moynihan:2020ejh,Johnson:2020pny,Plefka:2019wyg,Carrasco:2020ywq}.

On the other hand, in \cite{Penrose:1976jq} Penrose argued against perturbation theory. His seminal
work indeed showed that the Penrose transform is an exact formula for self-dual solutions \cite{Tod:1978,Tod:1979tt,Penrose:1976jq,PENROSE197765,Mason:1991rf,Mason:2005qu,Chacon:2020fmr}. Here we have relied on perturbation theory.
Still, there is hope that both approaches can be reconciled through
the connection with amplitudes. For instance, amplitudes in self-dual
gravity are known to be 1-loop exact \cite{Bern:1998xc}, which suggests a close connection
to the exact Penrose transform (see \cite{Ball:2021tmb,Adamo:2021dfg,Adamo:2021lrv,Adamo:2021zpw} for related recent works).

Acknowledgements: We thank Y.F. Bautista, E. Casali and A. Laddha for discussions. We thank J. Vines for sharing the note \cite{Vinesx} which explained relevant sections of \cite{penrose1984spinors}.
We also especially thank E. Crawley, N. Miller and A. Strominger for collaboration in \cite{Crawley}, which led to the ideas presented here. This work
was supported by DOE grant de-sc/000787 and the Black Hole Initiative at Harvard University,
which is funded by grants from the John Templeton Foundation and the Gordon and Betty Moore
Foundation. The author also receives support from the Harvard Society of Fellows.

\bibliography{references}
\bibliographystyle{aipnum4-1}

\newpage
\onecolumngrid
\section*{Appendices}
\appendix
\section{I. Twistors in $(1,3)$ Spacetime and Radiative Solutions}\label{ap:I}

In a beautiful geometrical construction, Penrose derived a formula
that enables direct bulk reconstruction of free radiative solutions \cite{penrose1984spinors}.
It is based on a map from twistors to null infinity. Before we revisit
it in this appendix, let us characterize radiative solutions in the somewhat more standard
fashion.

Continuing with the scalar model \eqref{eq:adv}, the purely radiative
solution is constructed by subtracting the retarded and advanced solutions,
such that Coulomb modes cancel out
\begin{align}
\phi(x) & :=\Phi_{R}-\Phi_{A}=\int\frac{d^{4}k}{(2\pi)^{3}}e^{ikx}i\left[\theta(k^{0})+\theta(-k^{0})\right]\delta(k^{2})\mathcal{J}(k)\,\,\,,\Phi_{A}=\Phi_{R}^{*}\label{eq:phirr}
\end{align}
This field fulfils $\partial^{2}\phi=0$ everywhere and hence can
be expanded in free modes. Radiation is given exclusively by the leading
$1/r$ component of the fields at future null infinity. In this regime,
$\phi(x)$ contains the same data as the retarded solution $\phi_{R}(x)$,
since the advanced solution vanishes. Thus \eqref{eq:phirr} confirms
the correspondence between radiative data and the on-shell modes $k^{2}=0$.
The presence of the step functions shows that we include both particle/antiparticle
phase space as expected for the retarded conditions as opposed to Feynman's. On the other
hand, for Coulomb solutions with $k^{0}=0$ we find that $\phi_{R}=\phi_{A}$
and hence the integral vanishes. 

We usually characterize null (radiation) data as follows. As $r\to\infty$, with
$u:=t-r$ fixed, the saddle point approximation can be used to localize
the oscillatory integrals in \eqref{eq:phirr}. They will only receive
contributions from the null directions $k_{\mu}=\omega\ell_{\mu}(x)$
that reach a particular point $x$, where
\begin{equation}
\ell_{\mu}(x)=\left(1,\hat{n}=\frac{\vec{x}}{r}\right)\,,\label{eq:lnk}
\end{equation}
parametrizes the celestial sphere. Using $u=\ell(x)\cdot x$ the radiative
data is then given by the asymptotic mode expansion \cite{Strominger:2017zoo}
\begin{equation}
\tilde{\phi}(u,\mathcal{\ell}):=\lim_{r\to\infty}r\Phi_{R}(x)=\frac{1}{8\pi^{2}}\int_{-\infty}^{\infty}d\omega\times e^{i\omega u}\mathcal{J}(\omega\ell^{\mu})\,.\label{eq:rdtg}
\end{equation}

We are now in a good position to connect to twistor theory. Indeed, a simple
way of resolving the bulk integration \eqref{eq:phirr} has been already
provided in the main text: Note that the delta function in the integrand
amounts to extract the residue at $\xi=0$ in \eqref{eq:ieps}, with
$c=R$. Defining the `curvature' $\psi(u):=\partial_u \tilde{\phi}$ the result is then, as stated
\begin{equation}
\phi(x)=\int_{\mathbb{CP}^{1}}\frac{\langle\lambda d\lambda\rangle[\tilde{\lambda}d\tilde{\lambda}]}{(2\pi)^{2}}\psi(\ell\cdot x,\ell)\,\,\,\,,\ell=|\lambda\rangle[\tilde{\lambda}|\label{eq:hkfs}
\end{equation}
(the field $\psi$ is the scalar analog of the Weyl radiative component $\psi_4=C_{u z uz}$ in Bondi coordinates). Up to zero modes of $\tilde{\phi}(u)$ that do not enter in the curvature, namely terms that survive
as $u\to\pm\infty$, this result completely reconstructs the radiative
field in the bulk (zero modes correspond to $\mathcal{J}(k)=\delta(k^{0})\times\textrm{constant}$; in such case the field $\phi$ vanishes but its radiative data \eqref{eq:rdtg}
does not). Indeed this is nothing but the Kirchoff-d'Adh\'emar formula introduced
by Penrose in \cite{penrose1984spinors,kirchoff}. Our derivation in momentum space allowed us
to make direct contact with the S-Matrix. In contrast, his proof followed
a completely different argument, using the following twistor interpretation:
In $(1,3)$ signature the celestial sphere coordinates $\lambda$
and $\tilde{\lambda}$ are complex conjugates. One can thus define
the map from $\mathbb{TP}$ to future null infinity $\mathcal{I}^{+}$
as
\begin{equation}
Z=(\lambda,\tilde{\mu})\to(u=\frac{1}{2}[\tilde{\mu}\tilde{\lambda}],\lambda,\tilde{\lambda})\in\mathcal{I}^{+}
\end{equation}
Via such map, the radiative data \eqref{eq:rdtg} becomes a function
in twistor space $\tilde{\phi}(Z,\bar{Z})$. The integrand in \eqref{eq:hkfs},
namely $\rho_{x}\partial_{u}\tilde{\phi}$, is not holomorphic in
contrast with the Penrose transform. However, it is still straightforward
that \eqref{eq:hkfs} satisfies
\begin{equation}
\partial^{A\dot{A}}\partial_{A\dot{A}}\phi(x)=0\,.
\end{equation}
Under the incidence relation \eqref{eq:incf} the retarded time $u=\frac{1}{2}[\tilde{\mu}\tilde{\lambda}]$
admits the geometrical interpretation of being the point where the
ray $\ell=|\lambda\rangle[\tilde{\lambda}$ emanating along the null
cone of $x$ intersects future null infinity. The celestial sphere
parametrized by $\lambda$,$\tilde{\lambda}$ then corresponds to
a cross section of such null cone. Penrose's proof relies on the observation
that the integral \eqref{eq:hkfs} is nicely independent of such cross section.

\section{II. Potential Modes from the Worldline EFT}

Equation \eqref{eq:m3mp} corresponds to a formula for three-point amplitudes associated to a scalar field. As discussed throughout the text, such potentials correspond to on-shell free modes of photons or gravitons. Here we delve into this correspondence for the case of gauge-theory, the graviton situation being analogous but notationally inconvenient. For illustration purposes we follow a complementary picture to that of the scattering amplitudes in \cite{Bautista:2019tdr,Chung:2019duq}, provided
by worldline EFT models \cite{Porto:2005ac,Levi:2015msa,Vines:2017hyw,Guevara:2020xjx}. 
We consider particles carrying both electric
and magnetic charges, the latter corresponding to the NUT parameter in the gravitational case \cite{Guevara:2020xjx}.

A compact object in Maxwell theory can be described by a classical
worldline action carrying an infinite tower of Wilson coefficients.
The interacting piece of the worldline action is 
\begin{equation}
S_{\textrm{int}}=\int(q_{0}A_{\mu}+i\tilde{q}_{0}A_{\mu}^{*})u^{\mu}d\tau+i\sum_{n=0}^{\infty}\frac{1}{(n+1)!}\int d\tau\left[q_{n+1}(ia\cdot\partial)^{n}F_{\mu\nu}u^{\mu}a^{\nu}+i\tilde{q}_{n+1}(ia\cdot\partial)^{n}F_{\mu\nu}^{*}u^{\mu}a^{\nu}\right]
\end{equation}
Here $F_{\mu\nu}^{*}$ is the Hodge dual of $F_{\mu\nu}$, which can
be written as a closed form $F_{\mu\nu}^{*}=2\partial_{[\mu}A_{\nu]}^{*}$.
This action is closely related to the one presented in \cite{Guevara:2020xjx} but
differs in that it carries magnetic charges. In fact, note that because
$a^{\mu}$ and $A_{\mu}^{*}$ are pseudovectors, a parity transformation
acts as $a^{\mu}\to-a^{\mu}\,\,A_{\mu}^{*}\to-A_{\mu}^{*}$, which
implies that the Wilson coefficients must have a parity odd piece, i.e. magnetic charges. At leading order in spin it is given by the coupling
$\tilde{q}_0$ in the first term. 

Now, since we are interested in vacuum configurations it is convenient
to introduce the self-dual components
\begin{equation}
A_{\mu}^{\pm}=\frac{1}{\sqrt{2}}(A_{\mu}\pm iA_{\mu}^{*})
\end{equation}
which parametrize the two degrees of freedom of a photon. They are
independent modes with associated currents
\begin{align}
j^{\pm\mu}(x)=\frac{\delta S_{\textrm{int}}}{\delta A_{\mu}^{\pm}(x)} & =\int d\tau\left[c_{0}^{\pm}u^{\mu}+\sum_{n}\frac{2c_{n+1}^{\pm}(ia\cdot\partial)^{n}}{(n+1)!}iu^{[\mu}a^{\nu]}\partial_{\nu}\right]\delta^{4}(x-x(\tau))\,\,\,,\label{eq:jxp}\\
c^\pm_{n} & =\frac{q_{n}\pm \tilde{q}_{n}}{\sqrt{2}}\,.\nonumber 
\end{align}
Assuming no incoming radiation, the trajectories of the compact object
are given by constant vectors $u^{\mu}(\tau)=u^{\mu}$ and $a^{\mu}(\tau)=a^{\mu}$,
hence we can take $x^{\mu}(\tau)=u^{\mu}\tau$. Further dropping the
total derivative term $u^{\mu}\partial_{\mu}=\frac{d}{d\tau}$ (the analog of three-point kinematics $u\cdot k$=0) a
short computation shows that in Fourier space \eqref{eq:jxp} becomes
\begin{equation}
j^{\pm\mu}(k)=u^{\mu}2\pi\delta(u\cdot k)\times\rho^{\pm}(\vec{k})\,\,,\textrm{with }\,\,\rho^{\pm}(\vec{k})=\sum_{n}\frac{c_{n}^{\pm}(\vec{a}\cdot\vec{k})^{n}}{n!}\label{eq:jpu}
\end{equation}
The charge densities $\rho^{\pm}$ then yield a multipole expansion
in momentum space, and indeed correspond to the three-point amplitudes
of the main text. To connect with the more familiar definition of
multipole moments we introduce position-space densities via
\begin{align}
\rho^{\pm}(k) & =\int d^{3}xe^{i\vec{k}\cdot\vec{x}}\rho^{\pm}(x)\nonumber \\
 & =\sum_{n}\frac{i^{n}\vec{k}_{i_{1}}\cdots\vec{k}_{i_{n}}}{n!}\int d^{3}x\vec{x}_{i_{1}}\cdots\vec{x}_{i_{n}}\rho^{\pm}(\vec{x})\nonumber \\
 & =:\sum_{n}\frac{i^{n}\vec{k}_{i_{1}}\cdots\vec{k}_{i_{n}}Q_{i_{1}\cdots i_{n}}^{\pm}}{n!}
\end{align}
where we have expanded in $\vec{k}$, after which we defined the multipole
moments of a charge distribution as usual. By comparing to \eqref{eq:jpu}
we obtain
\begin{equation}
Q_{i_{1}\cdots i_{n}}^{\pm}=(-i)^{n}c_{n}^{\pm}a_{i_{1}}\cdots a_{i_{n}}
\end{equation}
as in the main text. These multipole moments are transverse with respect
to $u^{\mu}$ but fail to be trace-free. However, this distinction
is irrelevant as we now show. 

As argued in the text, the two free degrees of freedom can be described
by self-dual potentials. This is because in a self-dual configuration
the magnetic and electric fields are proportional to $E_{i}^{\pm}=\partial_{i}\phi^{\pm}$ (or their `gravitoelectric' versions in the gravity case).
The potential follows from \eqref{eq:jpu}
\begin{align}
\phi^{\pm}(x) & =\int\frac{d^{4}k}{(2\pi)^{4}}\frac{2\pi\delta(u\cdot k)\times\rho^{\pm}(\vec{k})}{k^{2}}=\int\frac{d^{3}k}{(2\pi)^{3}}e^{i\vec{k}\cdot\vec{x}}\frac{\rho^{\pm}(\vec{k})}{\vec{k}^{2}}\nonumber \\
 & =\sum_{n}\frac{i^{n}Q^{\pm}_{i_{1}\cdots i_{n}}}{n!}\int\frac{d^{3}k}{(2\pi)^{3}}e^{i\vec{k}\cdot\vec{x}}\frac{\vec{k}_{i_{1}}\cdots\vec{k}_{i_{n}}}{\vec{k}^{2}}\label{eq:fdjg}
\end{align}
The strutures under the integral are now the $so(3)$ STF tensors, given by
(see next section for a spinorial analog)
\begin{equation}
\int\frac{d^{3}k}{(2\pi)^{3}}e^{i\vec{k}\cdot\vec{x}}\frac{\vec{k}_{i_{1}}\cdots\vec{k}_{i_{n}}}{\vec{k}^{2}}=\partial_{i_{1}}\cdots\partial_{i_{n}}\int\frac{d^{3}k}{(2\pi)^{3}}e^{i\vec{k}\cdot\vec{x}}\frac{1}{\vec{k}^{2}}=\partial_{i_{1}}\cdots\partial_{i_{n}}\frac{1}{4\pi r}=\frac{(2n-1)!!}{(-1)^{n}r^{n+1}}\hat{x}_{i_{1}}\cdots\hat{x}_{i_{n}}-\textrm{traces}\label{eq:refsd}
\end{equation}
They are STF because tracing two derivatives leads to $\partial^{2}1/r=0$
away from $r=0$. This entails that traces of $Q_{i_{1}\cdots i_{n}}$
in \eqref{eq:fdjg} will not contribute to the potential. In other
words, we are free to evaluate the current $\rho^{\pm}(\vec{k})$
by dropping contractions $\vec{k}^{2}$ in momentum space, since they
lead to ultra-local terms. Indeed, in $(2,2)$ signature
we can evaluate $\rho^{\pm}(\vec{k})$ at strict $\vec{k}^{2}=0$ with $\vec{k}\neq 0$,
where it becomes a three-point amplitude! This complexified interpretation of the potential was indeed the starting observation of \cite{Guevara:2017csg}.

Finally, we note that the physical interpretation of the tensors in
\eqref{eq:fdjg} follows by examining the ``minimal-coupling'' case
$c_{n}=1$,
\begin{equation}
\phi^{+}(x)=\sum_{n}\frac{i^{n}a_{i_{1}}\cdots a_{i_{n}}}{n!}\partial_{i_{1}}\cdots\partial_{i_{n}}\frac{1}{4\pi r}=\frac{1}{4\pi|\vec{x}+i\vec{a}|}
\end{equation}
which is the generating function of $m=0$ spherical harmonics if
$a$ is aligned with the $z$ axis. Thus the stationary multipole
moments appearing in the series \eqref{eq:fdjg} correspond to $m=0$
spherical harmonics. General $m$ is obtained from general, non-stationary,
multipoles as we now outline.

\section{III. Multipoles from Vectors to Twistors}

Throughout the text we use both vector and spinorial/twistor language
to describe multipoles. We outline here how to relate both forms.

In its most basic form multipoles can be defined as conjugates to
spherical harmonics. These form irreducible representations of the
little group associated that preserves certain time direction. Let
$u^{\mu}$ be such direction: We can thought it as a four-velocity
of a massive particle. A spin-$s$ representations is then given by
$2s+1$ polarizations
\begin{equation}
u_{\mu_{1}}\varepsilon_{m}^{\mu_{1}\cdots\mu_{s}}=\eta_{\mu_{1}\mu_{2}}\varepsilon_{m}^{\mu_{1}\cdots\mu_{s}}=0\,\,\,,\,-s\leq m\leq s\,.\label{eq:tran}
\end{equation}
i.e. by completely symmetric, trace-free (STF) tensors, yielding an
irreducible representation of the massive little group, i.e. $so(3)$
in $(1,3)$ signature. 

Consider now powers of the direction vector $\hat{x}^{\mu}=(0,\hat{x}(\theta,\phi))$
given by $\hat{x}_{\mu_{1}}\cdots\hat{x}_{\mu_{2}}$. Since these
products have non-vanishing traces they are not STF. Instead we can
extract their $2s+1$ irreducible components simply by projecting
with the complete set of states \eqref{eq:tran}, giving
\begin{equation}
Y_{sm}(\theta,\phi)=\varepsilon_{m}^{\mu_{1}\cdots\mu_{s}}\hat{x}_{\mu_{1}}\cdots\hat{x}_{\mu_{s}}\label{eq:ham}
\end{equation}
i.e. the standard spherical harmonics. The $so(3)$ action of polarization
tensors then acts naturally on these functions. 

Alternatively, a manifestly STF version of the tensors is obtained
from the generating function $1/r$ as in the last section, using
\eqref{eq:refsd} we can rewrite the harmonics as
\begin{equation}
Y_{sm}(\theta,\phi)=\frac{(-1)^{s}r^{s+1}}{(2s-1)!!}\times\varepsilon_{m}^{\mu_{1}\cdots\mu_{s}}\partial_{\mu_{1}}\cdots\partial_{\mu_{s}}1/r\label{eq:ysmg}
\end{equation}
In this form we can replace, for the case of stationary multipoles,
$\varepsilon_{m=0}^{\mu_{1}\ldots\mu_{s}}\to\hat{a}^{\mu_{1}}\cdots\hat{a}^{\mu_{s}}$
since all the traces are projected out. For general $m$ in the non-stationary
case, each of the polarization tensors \eqref{eq:tran} corresponds
to an independent multipole structure, i.e. functions on the sphere
can be expanded as
\begin{equation}
\sum_{sm}a_{sm}Y_{sm}(\theta,\phi)=\sum_{s}Q_{i_{1}\cdots i_{s}}\hat{x}^{i_{1}}\cdots\hat{x}^{i_{s}}\,\,\,,\,\,Q_{i_{1}\cdots i_{s}}:=\sum_{m}a_{sm}\varepsilon_{i_{1}\cdots i_{s}}^{m}
\end{equation}
Further note that due to the condition \eqref{eq:tran} we can replace
the direction vector by the null direction $\ell_{\mu}=(1,\hat{x})$
used in the main text, giving
\begin{equation}
Y_{sm}(\theta,\phi)=\varepsilon_{m}^{\mu_{1}\cdots\mu_{s}}\ell_{\mu_{1}}(\theta,\phi)\cdots\ell_{\mu_{2}}(\theta,\phi)\,.\label{eq:ellx}
\end{equation}
Let us now translate the discussion to spinors. Their advantage is
that STF tensors can be constructed trivially. Indeed, as exploited
in \cite{Arkani-Hamed:2017jhn}, representations of $su(2)\approx so(3)$ are completely
symmetric spinors rather than STF tensors. To see this, we use that
the momentum vector defines a canonical basis of spinors
\begin{equation}
u_{A\dot{A}}=\sigma_{0}=|1_{a}\rangle[1^{a}|=|1_{+}\rangle[1_{-}|-|1_{+}\rangle[1_{-}|\,\,\,
\end{equation}
where $a=1,2$ are little-group indices. Note that the orthogonal
vectors to $u^{\mu}$ are simply the three symmetric components $\varepsilon_{A\dot{A}}^{ab}=|1^{(a}\rangle[1^{b)}|$.
It is customary to take the spin direction as $a_{A\dot{A}}=a\varepsilon_{A\dot{A}}^{+-}=a\sigma_{z}$.
Now, the STF tensors \eqref{eq:tran} are simply, up to a normalization, 
\begin{equation}
\varepsilon_{A_{1}\dot{A}_{1}\cdots A_{s}\dot{A}_{s}}=|1^{(a_{1}}\rangle[1^{b_{1}}|\cdots|1^{a_{s}}\rangle[1^{b_{s})}|\,\,\,,\,a_{i},b_{j}\in\{0,1\}\label{eq:polfs}
\end{equation}
Crucially, this is not simply a tensor product of $\varepsilon_{A\dot{A}}^{ab}=|1^{(a}\rangle[1^{b)}|$
vectors: Due to the full symmetrizations of little-group indices,
one can check that the traces vanish by contracting both sides with
$\eta^{\mu_{i}\mu_{j}}\to\epsilon^{A_{i}A_{j}}\epsilon^{\dot{A}_{i}\dot{A}_{j}}$.
We may omit the explicit little-group indices hereafter. Further contracting
both sides with $u^{A\dot{A}}$ shows that they are indeed transverse
and hence fulfill \eqref{eq:tran}. Using the form \eqref{eq:polfs},
the harmonics \eqref{eq:ham} now read
\begin{align}
Y_{sm}(\theta,\phi) & =\frac{1}{2^{s}}\langle1|\hat{x}|1]\cdots\langle1|\hat{x}|1]\nonumber \\
 & =\frac{1}{2^{2s}\langle\lambda|u|\tilde{\lambda}]^{s}}\langle1\lambda\rangle[\tilde{\lambda}1]\cdots\langle1\lambda\rangle[\tilde{\lambda}1]\label{eq:ystw}
\end{align}
The second line is the spinor version of \eqref{eq:ellx}, where $\ell=\frac{1}{2}\frac{|\lambda\rangle[\tilde{\lambda}|}{\langle\lambda|u|\tilde{\lambda}]}$
parametrizes $\mathbb{C}P^{1}$. Now, using that $|1^{b_{1}}]=\tilde{u}|1^{b}\rangle$
and the definition \eqref{eq:key} for the spatial vector $\vec{x}=X_{AB}$, we
can write
\begin{equation}
Y_{sm}(\theta,\phi)=\frac{1}{(2r)^{s}}\langle1|X|1\rangle\cdots\langle1|X|1\rangle=\frac{1}{(2r)^{s}}\varepsilon^{A_{1}A_{2}\cdots A_{2s}}X_{A_{1}A_{2}}\cdots X_{A_{2s-1}A_{2s}}
\end{equation}
Following the twistor construction \cite{curtis} we defined the multipole
components as
\begin{equation}
\varepsilon_{A_{1}B_{1}A_{2}\cdots B_{s}}:=\varepsilon_{A_{1}\dot{A}_{1}\cdots A_{s}\dot{A}_{s}}u_{\,\,\,B_{1}}^{\dot{A}_{1}}\cdots u_{\,\,\,B_{s}}^{\dot{A}_{s}}
\end{equation}
Note that the particular case of $m=0$, corresponding to stationary
multipoles, is again obtained by replacement $\varepsilon_{A_{1}B_{1}A_{2}\cdots B_{s}}\to A_{(A_{1}B_{1}}\cdots A_{A_{s}B_{s})}$
as in the main text. Actually, symmetrization here is not needed if
we use the twistor version of \eqref{eq:ysmg} which we now derive.

Let us focus on $(2,2)$ signature, where the little group is $so(2,1)\approx sl(2,\mathbb{R})$. Just as the identity \eqref{eq:refsd}
leads to their momentum representation, the spherical harmonics also
fulfill an analogous identity leading to their twistor representation
\begin{align}
\int_{\mathcal{C}}\frac{\langle\eta d\eta\rangle}{2\pi}\frac{\eta_{A_{1}}\cdots\eta_{A_{2s}}}{(\eta^{A}X_{AB}\eta^{B})^{s+1}} & =\frac{(-1)^{s}}{s!}\frac{\partial}{\partial X^{A_{1}A_{2}}}\cdots\frac{\partial}{\partial X^{A_{2s-1}A_{2s}}}\int_{\mathcal{C}}\frac{\langle\eta d\eta\rangle}{2\pi}\frac{1}{\eta^{A}X_{AB}\eta^{B}}\label{eq:ide}\\
 & =\frac{(-1)^{s}}{s!}\frac{\partial}{\partial X^{A_{1}A_{2}}}\cdots\frac{\partial}{\partial X^{A_{2s-1}A_{2s}}}\left(\frac{1}{2\sqrt{|X|}}\right)\nonumber 
\end{align}
Also analogously to the momentum integral \eqref{eq:refsd}, it is
direct that this generates irreducible representations of the little
group. Indeed, one can alternatively use $sl(2,\mathbb{R})$ covariance
to guess that the result must be a completely symmetric tensor which
scales as $\sim1/X^{s+1}$, i.e.
\begin{equation}
\int_{\mathcal{C}}\frac{\langle\eta d\eta\rangle}{2\pi}\frac{\eta_{A_{1}}\cdots\eta_{A_{2s}}}{(\eta^{A}X_{AB}\eta^{B})^{s+1}}=\frac{(2s-1)!!}{s!}\times\frac{1}{\sqrt{2}}\frac{X_{(A_{1}A_{2}}\cdots X_{A_{2s-1}A_{2s})}}{(X_{AB}X^{AB})^{s+1/2}}\label{eq:exint}
\end{equation}
with $X_{AB}X^{AB}=2|X|$. The numerical prefactor is fixed by projection.
This holds for a general symmetric tensor $X_{AB}$. Note that using
\eqref{eq:ystw}, \eqref{eq:ide} and \eqref{eq:exint} we can write
the spinoral version of \eqref{eq:ysmg}: 
\begin{align}
Y_{sm}(\theta,\phi) & =\frac{(-1)^{s}r^{s+1}}{(2s-1)!!}\varepsilon^{A_{1}A_{2}\cdots A_{2s}}\frac{\partial}{\partial X^{A_{1}A_{2}}}\cdots\frac{\partial}{\partial X^{A_{2s-1}A_{2s}}}\left(\frac{1}{r}\right)\,.
\end{align}

\section{IV. Convergence and Relation to Previous Formulae}

An integral formulae was already given in \cite{Crawley} for stationary metric
perturbations and their curvature in $(2,2)$, starting from three-point
amplitudes. Convergence of the integration was analyzed only for the
Taub-NUT solution and played a key role in constructing its linearized
version in $(2,2)$ Klein space. 

The goal of this appendix is then two-fold. For one, we analyze the
convergence of our integral formulae for the most general stationary
solution. Second, we will recover the integral formulae of \cite{Crawley}
thereby completing its convergence analysis.

First consider the Penrose-Ward formula
\begin{align}
\phi & =:\int_{\mathbb{R}}\frac{\langle\eta d\eta\rangle}{2\pi}\rho_{x}\mathfrak{M}_3(\eta,\tilde{\eta})\,,\label{eq:phire}
\end{align}
for the most general stationary amplitude. Using \eqref{eq:fm2m3},
\eqref{eq:m3mp}, together with the definition \eqref{eq:key} this
is
\begin{equation}
\mathfrak{M}_3(\eta,\tilde{\eta})=\int_{0}^{\infty}\frac{d\omega}{4\pi}e^{-\frac{\omega}{2}[\tilde{\eta}|u|\eta\rangle}\times\sum_{n}\frac{c_{n}}{n!}\left(\frac{\langle\eta|A|\eta\rangle\omega}{2}\right)^{n}\,.\label{eq:f2n}
\end{equation}
That this is a well-defined Laplace transform is because
three-point amplitudes, as opposed to higher-point amplitudes, are
analytic functions of the energy $\omega$. As explained in \cite{Guevara:2018wpp}, such \textit{soft expansion} is in correspondence with the multipole expansion of the source. Because of this we assume, as in textbooks, that it is uniformly convergent in the long-wavelength regime. This is just to say that the integral can be evaluated term by term in the series.

Consider $(2,2)$ signature where the twistor $Z=(\eta,\tilde{\eta})$
is real valued. The integral defined in \eqref{eq:f2n} only converges
on a region of real twistor space, given by
\begin{equation}
[\tilde{\eta}|u|\eta\rangle=Z^{\alpha}\Sigma_{\alpha\beta}Z^{\beta}>0
\end{equation}
Through the incidence relation, this region corresponds to the following
region of flat space
\begin{equation}
\rho_{x}[Z^{\alpha}\Sigma_{\alpha\beta}Z^{\beta}]=\langle\eta|\tilde{x}u|\eta\rangle=\eta^{A}X_{AB}\eta^{B}>0\,\,\,\,\textrm{for all \ensuremath{\eta^{A}}\,.}\label{eq:form}
\end{equation}
The above implies $\det(X_{AB})=z^{2}-x^{2}-y^{2}>0$ which is the
so-called Rindler wedge. This extends the analysis done in \cite{Crawley}
for Taub-NUT to the case of general stationary solutions. Note that
the Laplace transform \eqref{eq:f2n} can also be analytically extended
to complex twistors $Z^{\alpha}$, in particular the rotation 

\begin{equation}
[\tilde{\eta}|\to e^{i\alpha}[\tilde{\eta}|\,\,\,\,0<\alpha<\pi/2\label{eq:rote}
\end{equation}
can be compensated by a deformation of the $\omega$ contour in the
complex plane. The end result is that as long as $\textrm{Re}[\tilde{\eta}|u|\eta\rangle>0$
(i.e. $\alpha<\pi/2$) we obtain
\begin{equation}
\mathfrak{M}_3(\eta,\tilde{\eta})=\sum_{n}\frac{c_{n}}{2\pi}\frac{\langle\eta|A|\eta\rangle^{n}}{[\tilde{\eta}|u|\eta\rangle^{n+1}}
\end{equation}
The continuation for \eqref{eq:rote} is the reason we could replace
$i[\tilde{\eta}|$ by $[\tilde{\eta}|$ in our original formula \eqref{eq:fm2}
(the original formula corresponds to $\alpha=\pi/2$, which requires
an $i\epsilon$ prescription). Note that through the incidence relation
this is the same as $ix\to x$, which preserves the $(2,2)$ signature.

We briefly comment on the convergence of the Penrose-Ward transform
\eqref{eq:phire}. Since the transform is projective in $\eta$, it
can be evaluated simply via contour deformation. The fact that the
real form \eqref{eq:form} is positive-valued automatically implies
that its two roots (i.e. the poles of the integrand) are complex and
conjugate. Hence hence the real contour $\mathcal{C}=\mathbb{R}$
in \eqref{eq:phire} splits them as we required. On the other hand,
when $\det(X_{AB})=0$ the form \eqref{eq:form} admits a real solution
in $\eta$. This effectively pinches the contour and generates a singularity
in \eqref{eq:phire} Thus we recover the conclusion pointed out in
\cite{Crawley}: Singularities arise lie at the so-called Rindler horizon $z=\sqrt{x^{2}+y^{2}}$.

We can now compare our integral formulae with the expressions given
in \cite{Crawley}. Let us consider for instance, the Gaussian formula
\begin{equation}
C_{\dot{A}\dot{B}\dot{C}\dot{D}}=\int\frac{d^{2}\tilde{\lambda}}{(4\pi)^{2}}e^{-\frac{1}{2}[\lambda|xu|\lambda]}\tilde{\lambda}_{\dot{A}}\tilde{\lambda}_{\dot{B}}\tilde{\lambda}_{\dot{C}}\tilde{\lambda}_{\dot{D}}M_{3}^{-}\left(\langle\lambda|=[\tilde{\lambda}|\tilde{u},|\tilde{\lambda}]\right)\,,\label{eq:gas}
\end{equation}
given there for the anti-self dual components of the curvature (we have adjusted the conventions via $\mathcal{M}_{3}^{\textrm{there}}=M_{3}^{\textrm{here}}/8\pi G$,
so that the three-point amplitudes agree with \eqref{eq:3ptke} for
Kerr). Given that the classical three-point amplitudes \eqref{eq:m3mp}
are functions of $k=|\tilde{\lambda]}[\tilde{\lambda}|u$, hence even
in $|\tilde{\lambda]}$, we can restrict the integration to $|\tilde{\lambda}]^{2}>0$
by including a factor of $2$. This makes contact with our Feynman prescription, as it means we can parametrize $|\tilde{\lambda]}=\sqrt{\omega}\langle\eta|u$
with $\omega>0$ and a real projective $|\eta\rangle=(1\,\,\,\eta)$.
The result is 
\begin{equation}
C_{\dot{A}\dot{B}\dot{C}\dot{D}}=\int_{\mathbb{R}}\frac{\langle\eta d\eta\rangle}{2\pi}u_{\dot{A}A}\eta^{A}\cdots u_{\dot{D}D}\eta^{D}\int_{0}^{\infty}\frac{d\omega}{8\pi}e^{-\frac{\omega}{2}\langle\eta|X|\eta\rangle}M{}_{3}^{+}\left(\sqrt{\omega}|\eta\rangle,\sqrt{\omega}\langle\eta|u\right)\,.
\end{equation}
It is then straightforward to check that this agrees with our results
\eqref{eq:gp2}-\eqref{eq:pr2}. Similar analysis can be done for
the potential functions $\phi^{\pm}$. The advantage of the Penrose
transform, over the Gaussian integration \eqref{eq:gas} presented
in \cite{Crawley}, is that the former can be used to also evaluate the linearized
metric $h_{A\dot{A}B\dot{B}}$. In such case, the polarization tensors
$\epsilon_{\mu}\epsilon_{\nu}\to\frac{\tilde{\lambda}_{\dot{A}}\tilde{\lambda}_{\dot{B}}\mu_{A}\mu_{B}}{\langle\mu|u|\tilde{\lambda}]^{2}}$
yield poles in the Gaussian integration, but can be resolved easily
in the Penrose contour integral. The result is the linearized metric
in Bondi gauge as in e.g. \cite{Adamo:2021zpw}.

% \bibliography{references}
% \bibliographystyle{aipnum4-1}

\end{document}